# Direct observation of a dynamical glass transition in a nanomagnetic artificial Hopfield network

Michael Saccone[1*], Francesco Caravelli[1], Kevin Hofhuis[2,3], Sergii Parchenko[2,3], Scott Dhuey[4], Armin Kleibert[5], Cristiano Nisoli[1], and Alan Farhan[3*]

[1]Center for Nonlinear Studies and Theoretical Division, Los Alamos National Laboratory, Los Alamos, New Mexico 87545, USA.
[2]Laboratory for Mesoscopic Systems, Department of Materials, ETH Zurich, 8093 Zurich, Switzerland.
[3]Laboratory for Multiscale Materials Experiments (LMX), Paul Scherrer Institute, 5232 Villigen PSI, Switzerland.
[4]Molecular Foundry, Lawrence Berkeley National Laboratory, One Cyclotron Road, CA 94720, USA.
[5]Swiss Light Source, Paul Scherrer Institute, 5232 Villigen PSI, Switzerland.

e-mail: msaccone@lanl.gov, alan.farhan@gmx.net

**Spin glasses, generally defined as disordered systems with randomized competing interactions that result in an extensively degenerate ground state[1,2], are a widely investigated complex system. Theoretical models describing spin glasses are broadly used in other complex systems, such as those describing brain function[3,4], error-correcting codes[5], or stock-market dynamics[6]. This wide interest in spin glasses provides strong motivation to generate an artificial spin glass within the framework of artificial spin ice systems[7-9]. Here, we present the first experimental realization of an artificial spin glass, consisting of dipolar coupled single-domain Ising-type nanomagnets arranged onto an interaction network that replicates the aspects of a Hopfield neural network[10]. Using cryogenic x-ray photoemission electron microscopy (XPEEM), we performed temperature dependent imaging of thermally driven moment fluctuations within these networks and observed characteristic features of a two-dimensional Ising spin glass. Specifically, the temperature dependence of the spin glass correlation function follows a power law trend predicted from theoretical models on two-dimensional spin glasses[11]. Furthermore, we observe clear signatures of the hard to observe rugged spin glass free**

**energy**[1] **in the form of sub-aging, out of equilibrium autocorrelations**[12] **and a transition from convergent to divergent dynamics**[1, 13]**.**

Recent advances in nanofabrication techniques opened a pathway to create artificial spin systems that exhibit geometrical frustration and allow direct real-space observations of magnetic configurations[9]. Artificial spin ice systems, comprising Ising-type nanomagnets lithographically arranged onto two-dimensional square[14, 15] and kagome[16, 17] geometries, emerged as prominent examples in recent years. Artificial spin ices exhibiting thermally induced moment fluctuations[16] paved the way for a whole new line of research where Ising-type nanomagnets are arranged onto novel two-dimensional magnetically frustrated geometries that exhibit a variety of emergent phenomena, ranging from emergent magnetic charge screening[18, 19], effective reduced[20] and elevated[7] dimensionalities, effective or mediate interaction patterns[21, 22], to emergent topological order[23]. Despite this long list of success stories of artificial spin ice systems, the realization of an artificial spin glass system remained elusive. True spin glass systems not only arise from random interactions and competition between ferro- and antiferromagnet order, but also possess distinct thermodynamic and dynamical traits[1, 2]. The main challenge remained to design arrays of nanomagnets with a dipolar interaction network that leads to spin glass behavior. For example, using a Gaussian-type disorder in arranging Ising-type nanomagnets onto a two-dimensional plane[8] determined that the type and degree of disorder are crucial to the realization of spin glasses when an ideal balance between ferro- and antiferromagnetic interactions is achieved. Despite balancing competition between ferro- and antiferromagnetic interactions in a randomized array of nanomagnets, a spin glass phase appears inaccessible at finite temperatures using this approach[8]. Using the concept of effective dimensionality in interacting networks[25] and theoretical predictions that a spin glass phase can only be stabilized at finite temperatures when a critical effective dimension of 2.52 is surpassed[26], it was shown that tree-like nanomagnetic patterns with elevated effective dimensionality can be a successful strategy to increase the effective dimension well above this critical value[7]. However, fabricating extended and quasi-infinite tree-like structures remains a currently unsurpassable challenge.

Here, we seek to realize an artificial spin glass implementing a proof-of-principle Hopfield neural network[10] (see methods), a model of associative memory mathematically equivalent to a spin glass, to guide the disorder of artificial spin systems. Conceptually, associative memory does not require a perfectly identical scenario to identify a memory. For example, most people can recognize a familiar face if it is partially obscured or an entire song from a low-quality recording. Hopfield networks are dynamical systems that evolve toward memories when their inputs are within a neighborhood of those memories. The memories of these networks correspond to ground states of a spin system and are robust to noise (see schematic in Figure 1a, b and Supplementary Figure 1). This robustness corresponds to a broad basin of attraction surrounding the spin glass ground state, allowing the system, in theory, to relax towards the ground state at non-zero temperatures.

We fabricated nanomagnetic Hopfield networks (see methods) consisting of permalloy ($Ni_{80}Fe_{20}$) Ising-type nanomagnets with lengths $L = 300$ nm, widths $W = 100$ nm, and thicknesses $d = 2.7$ nm (see Figure 1c). The dimension of the nanomagnets was chosen to ensure thermally driven moment reorientations occurring at the timescale of a few seconds to occur at a blocking temperature $T_B = 110$ K. Following sample fabrication, the sample was kept in vacuum at room temperature for several weeks to allow the Hopfield networks to relax towards equilibrium low-energy states[19, 27] before it was transferred into the photoemission electron microscope (PEEM) for magnetic imaging, employing x-ray magnetic circular dichroism (XMCD) at the Fe $L_3$ edge[28] (see methods). In PEEM, the sample was cooled down to 105 K (below the blocking point) and imaged, to observe the frozen-in low-energy state achieved after thermal annealing (see Figure 2a). Then, the sample was heated up to 120 K (above the blocking temperature), to start our real-time observations of thermal fluctuations and various temperatures (see Supplementary Movies 1 and 2).

As a first characterization step, we extracted the temperature dependent dimensionless magnetic susceptibility[8] $\frac{m^2}{k_B}\chi(T)$ (see methods) and plotted its inverse as a function of temperature (see Figure 3a). Note that these measures and those extracted from them result from the collective system dynamics and therefore differ from those of isolated permalloy. Fitting this temperature dependence to a Curie-Weiss law, $\frac{k_B}{m^2}\frac{1}{\chi(T)} = \frac{(T-T_C)}{A}$, (see green dashed line in Figure 3a) revealed a Curie temperature $T_C = 27.6 \pm 15.7$ K. This temperature is far below the blocking temperature $T_B = 100$ K of the patterned nanomagnets, comparable to results obtained from nanomagnetic arrays with a Gaussian disorder[8] and confirms that interactions are well randomized in the Hopfield networks because ferromagnetic order does not dominate as seen previously[8].

In bulk experimental spin glass, a comparison between field cooled and zero field cooled systems typically shows signatures of a spin glass. Here, we provide a more direct characterization of the thermodynamics to explore spin glass behavior in these artificial Hopfield networks. We extracted both the standard spin correlation function $[C'(r)]_{av}$ and the unbiased spin glass correlation function[29] $[C^{SG}(r)]_{av}$ (see methods) and plotted them as a function of distance (see Figure 3b). We then fit these correlation functions with a spatial decay function in the form of $e^{-\frac{r}{L(T)}}$ and $e^{-\frac{r}{L^{SG}(T)}}$, with $L(T)$ and $L^{SG}(T)$ being the temperature dependent standard (blue squares in Figure 3b) and spin glass correlation lengths (red asterisks in Figure 3b), respectively. Fitting the temperature dependence of these correlation lengths to a power law of the form $f(T) = B(T - T_c)^v$ (see blue and red dashed lines in Figure 3b), we calculate a standard critical exponent $v = 0.171 \pm 0.606$ and a spin glass critical exponent $v^{SG} = 3.86 \pm 1.2$. The latter values come close to the critical exponent $v^{SG} = 3.559$ predicted for a two-dimensional Ising spin glass[11], indicating that our artificial Hopfield networks are ordering towards a spin glass transition.

The dynamics of spin glass vary significantly when two factors are changed: whether the system is in or out of equilibrium and above or below the glass transition. To complicate matters further, evidence suggests that there is not simply a single, fixed glass transition[1,2]. Typically, but not always, there is a second, dynamical transition temperature. This usually exceeds the "static" critical temperature and is characterized by shifting peaks of AC susceptibility in experiments[1]. Higher frequency measurements tend to increase the temperature at which AC susceptibility peaks, which occurs in part because of an increasingly prominent "memory" of previous states resulting from the slow exploration of phase space. Computational studies observe this transition through how different initial states maintain a finite overlap with one another over time[30], settling into distinct regions in phase space. Others characterize this as a transition from high temperature chaotic dynamics to low temperature convergent dynamics. Here we employ an analysis of the system's autocorrelation function, its imperfect power law decay, and the Lyapunov exponent[13] (see methods), all as a function of temperature.

Signatures of the system's state may be found directly from the two-point autocorrelation function (see Methods). Both the general shape of the function and the critical exponent resulting from a power law fit can help categorize the system. The log-log plot of the autocorrelation functions (Fig. 4a) all decrease in slope over time, indicating a variable critical exponent and, by extension, that the system has not yet relaxed to equilibrium. The critical exponent itself (Fig. 4b) reinforces this conclusion, as it is significantly lower than the minimum values predicted in equilibrium, $\nu = 0.395$[31] or $\nu = 0.5$[2]. Notably, non-equilibrium autocorrelations are often flatter with lower time elapsed[1].

To extract more information about the chaoticity of the system, we studied the Lyapunov exponent from the spin dynamics. Transitions from convergent to chaotic behavior begin when similar trajectories through phase space diverge exponentially and continue to diverge despite the phase space being bound. The time rate of the exponential behavior, the Lyapunov exponent, is positive when the system is divergent, potentially chaotic, and negative when the system is convergent. Using a data driven method[34], we find similar initial paths and use their average distance over time to extract the Lyapunov exponent for each temperature (see methods). The exponents transition from negative values at low temperatures to positive values at high temperatures (Fig. 4c), consistent with a dynamical transition.

Assessing the system's statics holistically, the dominance of the spin glass correlation length over the standard correlation length and its temperature dependence are hallmarks of a system with a glass ground state. Despite the system ordering as indicated by the increasing magnetic susceptibility with decreasing temperature (Fig. 3a), the standard correlation lengths (Fig. 3b, blue squares) are essentially noise. The power law fit determines that $\nu = 0.171 \pm 0.606$, confirming that the standard correlation function can no longer determine the order parameter. On the other hand, the spin glass correlation length grows rapidly as the system is cooled (Fig. 3b, red asterisks). Its power law fit produces a critical exponent of $\nu = 3.86 \pm 1.27$ which, despite the relatively large uncertainty, is only 8.52% from the theoretically known value for a two-dimensional spin glass, $\nu = 3.559 \pm 0.025$ [11]. The direct computation of the critical exponent in a physical system validates core components of spin glass theory without the need to rely on bulk measurements. Future experiments with longer time sequences and more temperatures can improve the statistics to further validate this critical scaling.

A dynamical analysis indicates a non-equilibrium temporal correlation and a dynamical transition support the hypothesis of a rough free energy landscape. The exact temperature dependence is non-universal, but the autocorrelation function of many spin glasses decays with a power law with an exponent ν = 0.5 at the Alameda-Thouless line in equilibrium[2] and ν = 0.395 for the Edwards-Anderson model[31]. This exponent typically decreases as temperature approaches zero. Experimental results and out of equilibrium simulations find that the critical exponent varies over long periods of spin glass aging[1]. It is common for $v$ to start small as the system initially explores the phase space (sub-aging) and then increase as a path towards lower energy states is found (aging)[12]. Careful observation of autocorrelation functions from our system shows that they tend to have a constant value of the exponent $v$ until the end of our measurements where $v$ begins to increase (Fig. 4a). The exponent $v$ varies earlier when temperature is higher. Combining this observation with the fact that all $v$ values (Fig. 4b) are far below anything predicted by equilibrium theory suggests that the system is out of equilibrium at all temperatures and relaxing in the sub-aging regime. Further, the faster relaxation of the higher temperature systems allows the systems to leave the sub-aging regime faster, resulting in more variable slopes and increasing the $v$ determined by the fit as it increases over time. Aside from this continuous evolution towards faster relaxation from sub-aging, there is another prominent trend in the autocorrelation functions. At 157 K and above, the values of the autocorrelation function remain relatively similar despite a decreasing slope. However, as the temperature drops between 157 K and 147 K, systems more rapidly diminish in their average autocorrelation, then slowly increase in their average autocorrelation after this initial decrease in temperature. The secondary increase in autocorrelation is likely due to lower fluctuation rates of the magnetic moments, but the initial dramatic decrease between 157 K and 147 K seems to arise from a dynamic transition. The Lyapunov exponents and the rough free energy landscape[33] of spin glasses further solidify this conclusion.

The Lyapunov exponents increase with increasing temperature (Fig. 4c), showing a tendency for similar initial states to diverge as the system heats up. The system transitions from convergent dynamics ($\lambda < 0$) to divergent dynamics ($\lambda > 0$) around 157 K, the same temperature where average autocorrelation jumps dramatically. This is consistent with the system settling into deeper free energy minima, after losing enough energy to no longer traverse a broader section of phase space (See Supplementary Movie 3), increasing the rate of relaxation and grouping together similar trajectories in the same basin. A grouping of trajectories explains the energetic origins of both dynamic transition and memory in spin glass, especially considering that the basin is likely centered around a state encoded into the underlying Hopfield network. As a whole, a varying relaxation over time and a rough free energy landscape are both hallmarks of a spin glass.

As annealing for these systems is further improved, direct real-space studies and investigation of the spin glass ground state will be accessible and assist in our understanding of equivalent NP hard problems[30] and brain science models[3]. The freedom to lithographically tweak these systems' interaction networks will allow for the representation of other computing problems. It has already been shown that nanomagnetic systems may potentially approach the Landauer limit at room temperature[34] and thus make excellent candidates for low energy computing. Altogether, visual access to the dynamics of these systems allows for a rich tool in comprehending novel disordered phases and the myriad of systems with glassy behavior.

## Methods

**Designing a nanomagnetic Hopfield network.** Both Hopfield networks and Ising spin systems evolve as governed by their "interaction" networks. Here we describe how those networks are defined and how they may be modified computationally to match one another prior to fabrication.

A Hopfield neural network of size $N$ is represented by a vector $S_i'^m$ of binary states (-1 and 1) at iteration $m$. A connectivity matrix $w_{ij}$ governs its dynamics via the rule

$$S_i'^{m+1} = f\left(\sum_{j=1}^{N} w_{ij} S_j'^m\right).$$

To maintain a binary range, the activation function is defined as $f(x) = \frac{|x|}{x}$, the "sign" function. The connectivity matrix is created from a set of $n$ patterns, $\xi_i^\nu$, each labelled by $\nu$, that $S_i'^m$ intends to "recall," or grow closer to, over several iterations. The storage is encoded in the connection or weight matrix by the Hebbian learning rule:

$$w_{ij} = \frac{1}{n}\sum_{\nu=1}^{n} \xi_i^\nu \xi_j^\nu.$$

Here we consider $\xi_i^\nu$ to be a random vector whose entries are independently drawn from the probability distribution $p(\xi_i^\nu = 1) = p(\xi_i^\nu = -1) = 0.5$. Practically speaking, patterns of interest may not take on this form, but a mapping of all bits from a set of patterns $\tilde{\xi}_i^\nu$ onto $\xi_i^\nu$ is possible if $n < N$. The attractors of this dynamical system are $\xi_i^\nu$, making them the "memorized" patterns of the system.

Hopfield showed that the iterative evolution described in equation (1) always decreases an effective Hamiltonian,

$$H_{eff} = -\frac{1}{2}\sum_{ij} w_{ij} S_i' S_j'.$$

The Ising Hamiltonian in zero field has the same form,

$$H_I = -\frac{1}{2}\sum_{ij} J_{ij} S_i S_j.$$

$J_{ij}$ is here determined by magnetic interactions and is analogous to the connectivity matrix $w_{ij}$. In artificial nanomagnets, dipolar interaction strength is determined by the distribution of magnetization and positions and orientations of the nanomagnets for patterned nanomagnetic systems and $S_i$ is the binary Ising variable indicating the orientation of the magnetization. To model the exact interaction strength for a collection of nanomagnets with positions $r_i$ and orientations $\theta_i$, we implement the compass needle model:

$$J_{ij} = -\left(\frac{1}{|\mathbf{r}_{ai} - \mathbf{r}_{aj}|} - \frac{1}{|\mathbf{r}_{ai} - \mathbf{r}_{bj}|} - \frac{1}{|\mathbf{r}_{bi} - \mathbf{r}_{aj}|} + \frac{1}{|\mathbf{r}_{bi} - \mathbf{r}_{bj}|}\right).$$

Since this model assumes interactions occur between magnetic charges at the ends of the nanomagnets, $\mathbf{r}_{ai}$ and $\mathbf{r}_{bi}$ are the positions of the positive and negative charge belonging to spin $i$ as determined by the lengths, positions, and orientations of the magnets.

To fabricate an Ising system equivalent to a Hopfield network, one must first reduce the difference between $w_{ij}$ and $J_{ij}$ as much as possible (see Fig. 1a-b). The scale of each is irrelevant, so they are both normalized by the average absolute interaction strength per neuron or spin. Specifically, $w'_{ij} = w_{ij}N/\sum_{ij}|w_{ij}|$ and $J'_{ij} = J_{ij}N/\sum_{ij}|J_{ij}|$. We then use machine learning methods to change $\xi_i^\nu$ and the positions and angles of a nanomagnetic design to minimize a cost function $C = \sum_{ij}(w'_{ij} - J'_{ij})^2$. We determine this through gradient descent of the continuous variables, $\mathbf{r}_i$ and $\theta_i$, and relatively quickly reach local minima. The cost function may be further reduced through modification of the discrete pattern states, $\xi_i^\nu$, as any states of interest may be mapped onto arbitrary stored patterns. A Monte Carlo Metropolis annealing of "energy" $C$ with "spins" $\xi_i^\nu$ is an appropriate method of further reducing the cost function. This is carried out with parallel tempering at 100 separate temperatures, the lowest temperature of which is used for the new $\xi_i^\nu$. The overall process of matching the systems progresses by alternating gradient descent and annealing until $C$ converges. The positions and orientations are then used to fabricate our nanomagnetic system (Fig. 1c).

**Sample fabrication.** Lift-off assisted electron beam lithography was used to generate nanomagnetic Hopfield networks. A 1×1 cm$^2$ Silicon (100) substrate is first spin-coated with a 70 nm thick layer of polymethylmethacrylate (PMMA) resist. Then, a VISTEC VB300 e-beam writer is used to define the Hopfield patterns onto the substrate. Following development of the exposed resist layer, a 2.7 nm thin Permalloy (Ni$_{80}$Fe$_{20}$) film is deposited on the substrate at a base pressure of 1.4×10$^{-7}$ torr, together with an Aluminum capping layer of 2 nm, to avoid fast oxidation. Then, the substrate is placed into Acetone for lift-off. The resulting nanomagnetic artificial Hopfield networks consisted of nanomagnets with lengths $L = 300$ nm and widths $W = 100$ nm.

**Photoemission electron microscopy (PEEM).** Magnetic imaging was performed at the PEEM endstation of the SIM beamline at the Swiss Light Source (SLS) and the 21-ID-2 Beamline at the National Synchrotron Light Source (NSLS), employing x-ray magnetic circular dichroism (XMCD) at the Fe L3 edge[28]. An XMCD image is a result of pixelwise division of images obtained with circular left and circular right polarized light. The typical dark and bright contrast is a direct measure of orientation of a magnetic moment with the incoming x-ray propagation vector. Moments with a non-zero component towards the incoming x-rays will appear dark, while moment pointing in the opposite direction will appear bright (Fig. 2a). 70 XMCD images were recorded every 14 seconds at 120 K, 130 K, 147 K, 157 K, 168 K, 181 K, and 196 K. Systemwide time evolution occurred on the order of seconds as indicated by Fig. 2b.

**Spin-spin correlations and magnetic susceptibility.** Temperature dependent spatial spin correlations are extracted using our previously employed method[8]. The spatial correlation function was calculated:

$$C(\boldsymbol{r}_{ij}) = \langle S_i S_j \rangle_T$$

where $S_i = \pm 1$ to represent the Ising state of spin $i$, $\boldsymbol{r}_{ij}$ is the distance between spins $i$ and $j$, and $\langle \cdots \rangle_T$ denotes a thermal average. The absolute value of this, $C'(\boldsymbol{r}_{ij}) = |C(\boldsymbol{r}_{ij})|$, was used for correlation function calculations. All correlation function values corresponding to $r - \Delta/2 < r_{ij} < r + \Delta/2$ where $\Delta$ is the distance between consecutive $r_k$, were averaged to a single value,

$$[C'(r)]_{av} = \frac{1}{N_{pair}} \sum_{ij} C'(\mathbf{r}_{ij}).$$

The decay of the correlation function is expected to follow an exponential function $[C'(r)]_{av} = e^{-\frac{r}{L(T)}}$, where $L(T)$ is the standard correlation length, which can also be plotted as function of temperature.

The dimensionless magnetic susceptibility $\chi$ was calculated from this correlation using the fluctuation dissipation theorem. This susceptibility $\chi$ was returned to appropriate dimensions by an additional factor, $m$, the magnetic moment of a single Ising macrospin:

$$\chi = \frac{m}{k_B T} \sum_{ij} C(\mathbf{r}_{ij}).$$

For the nanomagnets discussed here, the magnetic moment $m$ is calculated from a saturation magnetization, $M = 85$ kA/m found for similarly thin-film permalloy structures[16], to be $m = 5.41 \times 10^{-18}$ Am$^2$.

**Unbiased spin glass spin-spin correlation and correlation length.** The measurements determine $C_{ij}$, an estimate for the spin-spin correlation $\langle S_i S_j \rangle$, between all pairs of spins $i$ and $j$. Naturally, there is an uncertainty in the experimental results. Let us write

$$C_{ij} = \langle S_i S_j \rangle + \epsilon_{ij}. \quad (1)$$

Assuming that the system is equilibrated, the error $\epsilon_{ij}$ is a random variable with zero mean. We say that $C_{ij}$ is an unbiased estimator for $\langle S_i S_j \rangle$. (Unbiased means that if one repeats the set of measurements many times, then the average gets arbitrarily close to the exact answer). However, for spin glass correlations we need the square of the correlation function. In this case, $C_{ij}^2$ is a biased estimator for the spin glass correlation function $\langle S_i S_j \rangle^2$ because

$$C_{ij}^2 = \langle S_i S_j \rangle^2 + 2\langle S_i S_j \rangle \epsilon_{ij} + \epsilon_{ij}^2$$

and the term $\epsilon_{ij}^2$ has a non-zero mean. As a simple example, suppose that $C_{ij}$ is obtained from just one spin configuration. Then $C_{ij}^2 = \langle S_i S_j \rangle^2 = 1$, for all pairs. Hence summing over all pairs to give the spin glass susceptibility gives a completely wrong result. However, even if $C_{ij}$ is obtained from just one spin configuration, summing $C_{ij}$ over all pairs to get the ferromagnetic susceptibility gives a result which, though having quite large error bars, is nonetheless unbiased. From spin configurations one can calculate:

$\langle S_i S_j \rangle$ estimated from $C_{ij} = \frac{1}{n_t} \sum_{\alpha=1}^{n_t} S_i(t_\alpha) S_j(t_\alpha)$.

$\langle S_i S_j \rangle^2$ estimated from $C_{ij}^2 = \left[ \frac{1}{n_t} \sum_{\alpha=1}^{n_t} S_i(t_\alpha) S_j(t_\alpha) \right]^2$.

One can eliminate the bias in the above estimate for $\langle S_i S_j \rangle^2$ by dividing the $n_t$ measurement times into two equal halves, and correlating the spin product $\langle S_i S_j \rangle$ at a time $t_\alpha$ in the first half with the same spin product at the corresponding time in the second half, i.e.

estimate $\langle S_i S_j \rangle^2$ from $C_{ij}^{SG} = \frac{1}{n_t/2} \sum_{\alpha=1}^{n_t/2} S_i(t_\alpha) S_j(t_\alpha) S_i(t_{\alpha+n_t/2}) S_j(t_{\alpha+n_2/2})$.

If $t_{n_t}$ is greater than the relaxation time of the spins there is no correlation between the spins at the earlier and later times and so, on average, this expression for $C_{ij}^{SG}$ is equal to the desired quantity $\langle S_i S_j \rangle^2$ without any bias. This estimate of $C_{ij}^{SG}$ was mapped to $C^{SG}(r_{ij})$ for the nanomagnets and was spatially averaged to extract a spin glass correlation length:

$$[C^{SG}(r)]_{av} = \frac{1}{N_{pair}} \sum_{ij} C_{ij}^{SG}(r_{ij}).$$

**Dynamical Analysis**

Now, we turn our attention to temperature-dependent observations of thermal fluctuations in our artificial Ising spin glass structures. Spin glass is observed and theorized to relax in different ways depending on whether the spin is equilibrated and below the critical temperature[1, 2]. The equilibrium behavior of more ideal spin glasses is thoroughly catalogued through early theories on spin glasses. The two-point autocorrelation function,

$$C(t_1, t_2) = \langle S_j(t_2) S_j(t_1) \rangle_j,$$

where $\langle ... \rangle_j$ is an average over all spin indices, relaxes in many different forms depending on the whether or not the system is above the Alameda-Thouless line, a dynamical transition temperature depending on external field, and the precise model being used[2]. A power law relaxation[29] best fit our system:

$$C(t_1, t_2) = (t_2 - t_1)^{v(T)}.$$

v($T$) is an exponent that varies with temperature. The two-point autocorrelation function is extracted from our time dependent data at each temperature and fit to the power law decay to extract v($T$) (Fig 4. a-b).

One can help assess whether a dynamic series is chaotic through computation of the Lyapunov exponent. The notion of the Lyapunov exponent considers a system and a near identical duplicate with a small offset in initial conditions. The systems evolve in parallel and an appropriate measure of distance, $D(t)$, between the two systems is analyzed. If a system is chaotic, $D(t)$ should grow exponentially. If not, $D(t)$ will diminish. This mathematical form, $D(t) = D_0 e^{\lambda t}$, is fit by a variety of methods to determine the sign of $\lambda$, deemed the Lyapunov exponent, a positive value indicating chaos. Note that the metric $D(t)$ may exist in higher dimensions and yield a variety of Lyapunov exponents, a Lyapunov spectrum, where the sign of the largest exponent is used to evaluate whether or not the system is chaotic. The method employed here uses one average measure of distance to estimate this largest Lyapunov exponent.

We begin by considering a series of spin data, $S_j(t_i)$, as a dynamical sequence. The sequence is processed as follows[34]:

1. The mean period, $T$, of the system is estimated from the peak of the power spectrum of the sum over all spins. That is, from the spectrum $P(f) = \left| \mathcal{F}\left( \sum_{k=1}^{N} S_k(t_i) \right) \right|^2$, where

$\mathcal{F}(\cdots)$ is the 1-D Fourier transform, one can determine the frequency that corresponds to the maximum power, $f_{max}$, and then $T = 1/f_{max}$.

2. For each time $t_i$, this mean period was used to find a "nearest neighbor" state at time $\hat{t}_i$. That is, $\hat{t}_i$ is the time where $\sum_k \left(S_k(t_j) - S_k(\hat{t}_j)\right)^2$ is minimized under the condition that $\left|t_j - \hat{t}_j\right| > T$ to prevent just picking a temporally correlated state. This comparison between similar, but temporally disparate states is assumed to be close to an experiment where two separate states with similar initial conditions are evolved in parallel.

3. Compute the distances, $d_j(t_i) = \sum_k \left(S_k(t_j) - S_k(\hat{t}_j + t_i)\right)^2$, between these two states over all possible times (that is, cease computation when either $t_j$ or $\hat{t}_j + t_i$ grows beyond the size of the data set).

4. Average the logarithm of the distances over every starting point, $j$: $y(t_i) = \frac{1}{\Delta t}\left\langle \log\left(d_j(t_i)\right)\right\rangle_j$. Given the base hypothesis that similar states exponentially grow or shrink in distance, the slope of the linear fit to this data is the estimate of the largest Lyapunov exponent, $\lambda$.

These exponents were calculated at every temperature and plotted in Fig. 4g. The exponents increase with temperature, transitioning from negative to positive values between 157 K and 168 K. This indicates a leap into chaotic or divergent behavior.

## Acknowledgements

The authors would like to thank A.P. Young for fruitful discussions on the spin glass correlation function. A.F. and K.H. acknowledge support from the Swiss National Science Foundation (Projects No. 174306 and 172774, respectively). Funding was also received from the European Union's Horizon 2020 research and innovation program under Grant Agreement

## Contributions

M.S. and A.F. conceived the project. A.F. designed and performed the experiments with support from K.H., S.P. and A.K. S.D. fabricated the samples. M.S. analyzed and interpreted the data with support from F.C. and A.F. M.S. and A.F. wrote the manuscript with input from all other authors. C. N. and A. F. supervised the project.

## Competing Interests

The authors declare no competing financial interests.

## Data availability

The data that support the findings in this study are available from the authors upon request.

**Figures**

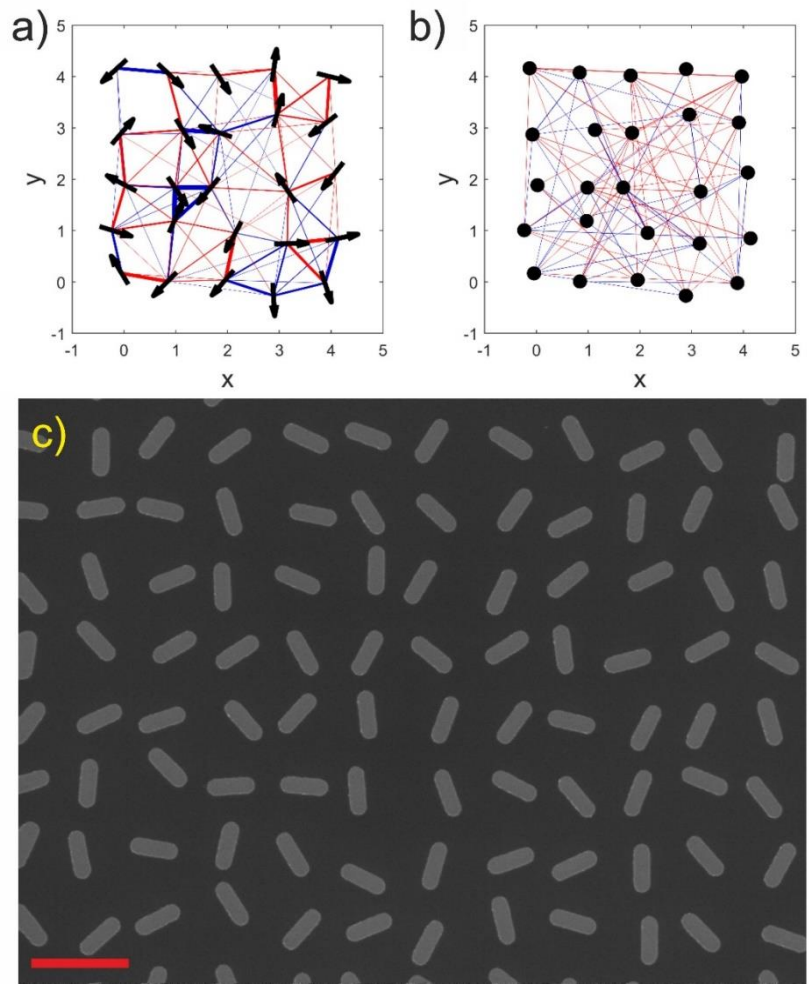

**Figure 1. Nanomagnetic artificial Hopfield networks. a,** An artificial spin glass with the coupling between spins represented as red ($J_{ij} > 0$ or ferromagnetic) and blue ($J_{ij} < 0$ or antiferromagnetic) lines, their thickness proportional to strength. **b,** A Hopfield neural network to which the spin glass was matched. The dots represent the neurons and the lines the dominant connections, drawn red if $w_{ij} < 0$ and blue if $w_{ij} > 0$. Further optimization will correct for mismatched features of these graphs, such as the higher proportion of non-local interactions in (b) and larger number of extraordinarily strong interactions in (a). **c,** Scanning electron microscopy (SEM) image of a portion of an artificial Hopfield network consisting of Ising-type nanomagnets with lengths $L = 300$ nm, widths $W = 100$ nm and thickness $d = 2.7$ nm. The red scale bar indicates a length of 600 nm.

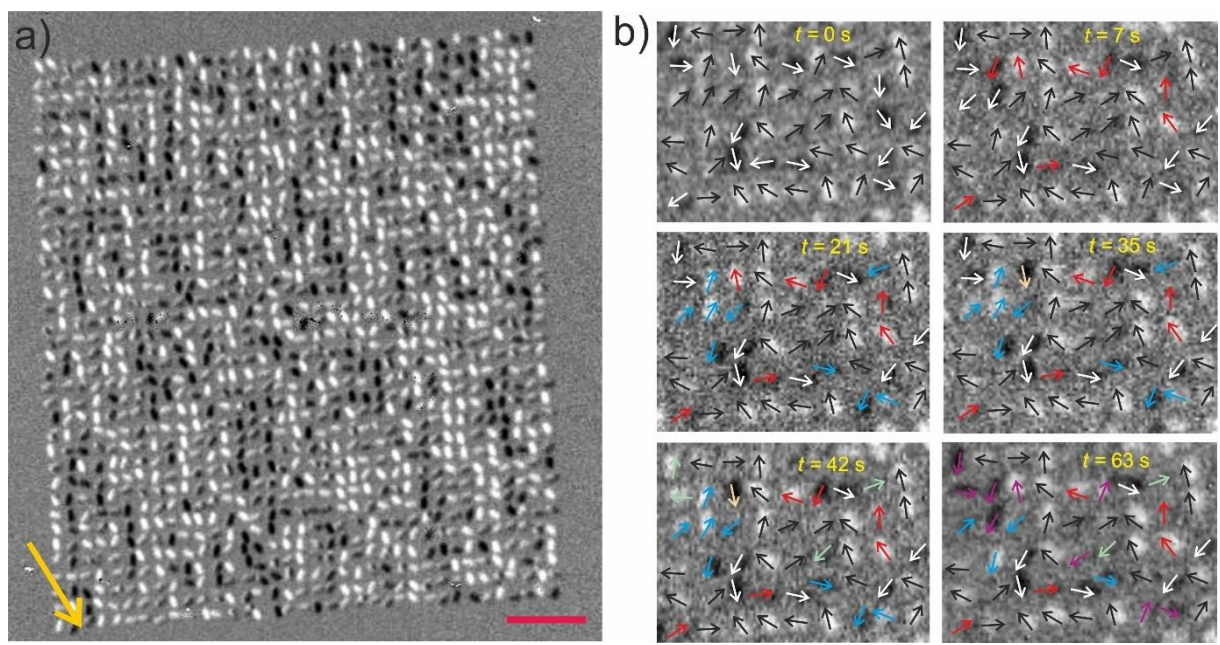

**Figure 2. Imaging low-energy moment configurations in a nanomagnetic artificial Hopfield network. a,** XMCD image recorded at 105 K of a frozen-in low-energy state achieved after thermal annealing. The yellow arrow indicates the direction of the incoming x-rays. Moments pointing the incoming x-rays appear dark, while moments in the opposite direction will appear bright. The red scale bar indicates a length of 2μm. **b**, Cropped XMCD image sequence of 6 images covering a timeframe of 63 seconds with moment reorientations occurring from frame to frame indicated with different colors.

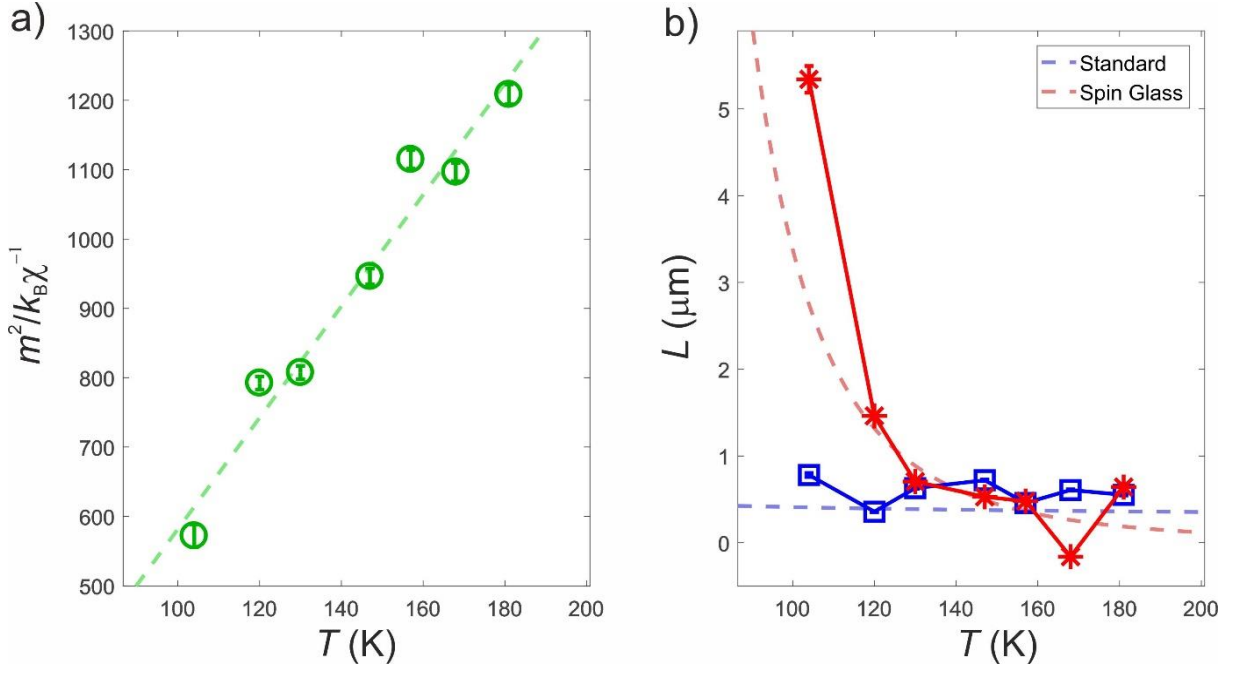

**Figure 3. Temperature dependent inverse susceptibility and correlation length derived from real-space observations. a,** The dimensionless, inverse susceptibility, $\frac{m^2}{k_B}\chi^{-1}$, of the annealed spin glass is plotted as green circles and fit to the Curie law, $\frac{m^2}{k_B}\chi(T) = \frac{A}{T-T_c}$, green dashed line, yielding $T_c = 27.6 \pm 15.7$ K and $A = -221$. **b,** The standard and spin glass correlation lengths extracted from their respective correlation functions (see Supplementary Figure 2) are plotted as blue squares and red asterisks, respectively. Fitting their critical behavior, $f(T) = B(T - T_c)^\nu$, finds a standard exponent and temperature of $\nu = 0.171 \pm 0.606$ and $B = 1.33 \frac{\mu m}{K^\nu}$, and spin glass parameters of $\nu = 3.86 \pm 1.27$ and $B = 1.78 \times 10^8 \frac{\mu m}{K^\nu}$.

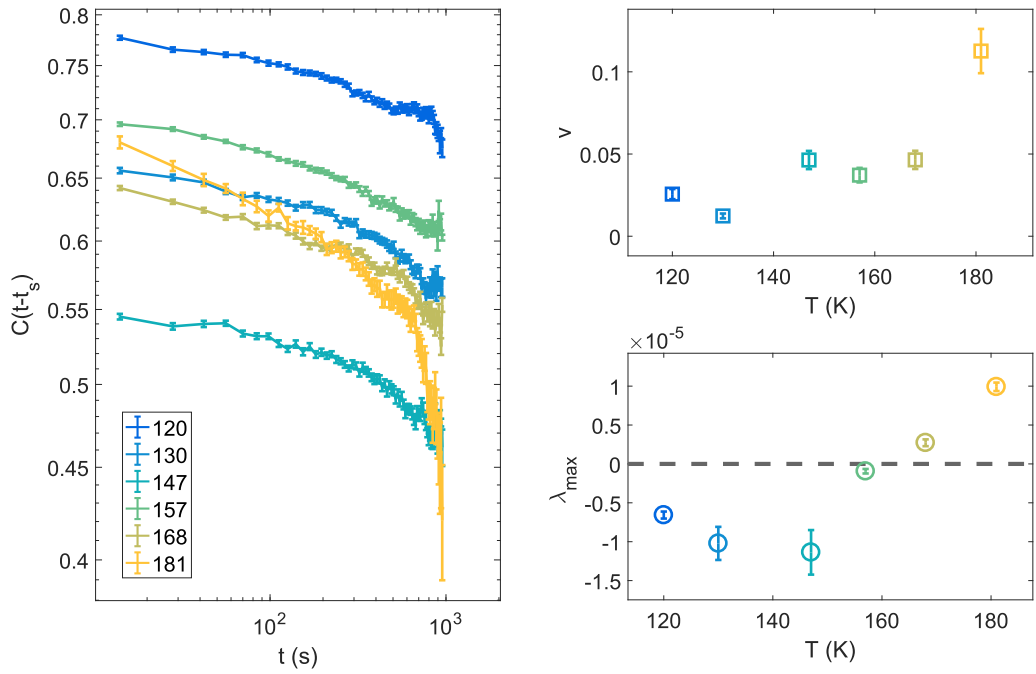

**Figure 4. Dynamical behavior of a nanoamgnetic Hopfield network. a,** The autocorrelation function plotted on a log-log plot for all lattice temperatures. If the anticipated power law decay is observed, the plots will be linear. **b,** The decay power $v$ fit from the autocorrelation function of the form $C(t - t_p) = C_0(t - t_p)^{-v}$. **c,** The Lyapunov exponent of moment dynamics plotted versus temperature.